\documentclass[onefignum,onetabnum]{siamart190516}

\usepackage[english]{babel}

\usepackage{amsmath}
\usepackage{mathtools}
\usepackage{amssymb}
\usepackage{listings}
\usepackage{float}
\lstdefinelanguage{Julia}%
  {morekeywords={abstract,break,case,catch,const,continue,do,else,elseif,%
      end,export,false,for,function,immutable,import,importall,if,in,%
      macro,module,otherwise,quote,return,switch,true,try,type,typealias,%
      using,while},%
   sensitive=true,%
   alsoother={$},%
   morecomment=[l]\#,%
   morecomment=[n]{\#=}{=\#},%
   morestring=[s]{"}{"},%
   morestring=[m]{'}{'},%
}[keywords,comments,strings]%

\usepackage{lipsum}
\usepackage{amsfonts}
\usepackage{graphicx}
\usepackage{epstopdf}
\usepackage{algorithmic}
\ifpdf
  \DeclareGraphicsExtensions{.eps,.pdf,.png,.jpg}
\else
  \DeclareGraphicsExtensions{.eps}
\fi

\newsiamremark{remark}{Remark}
\newsiamremark{hypothesis}{Hypothesis}
\crefname{hypothesis}{Hypothesis}{Hypotheses}
\newsiamthm{claim}{Claim}

\headers{Robust Numerical Solvers for Nuclear Fusion Simulations}{A. Quinlan V. Dwarka I. Holod M. Hoelzl}

\title{Towards Robust Solvers for Nuclear Fusion Simulations Using JOREK: A Numerical Analysis Perspective}

\author{Alex Quinlan \thanks{Department of Applied Mathematics, Delft University of Technology, Delft, the Netherlands
  (\email{alexsquinlan@gmail.com, v.n.s.r.dwarka@tudelft.nl})}
\and Vandana Dwarka \footnotemark[1] 
\and Ihor Holod \thanks{Max Planck Institute of Plasma Physics, Garching, Germany  (\email{ihor.holod@ipp.mpg.de, matthias.hoelzl@ipp.mpg.de})}
\and Matthias Hoelzl \footnotemark[2] }

\usepackage{amsopn}

\ifpdf
\hypersetup{
  pdftitle={Towards Robust Numerical Solvers for Nuclear Fusion Simulations Using JOREK: A Numerical Analysis Perspective},
  pdfauthor={Alex Quinlan Vandana Dwarka Ihor Holod Matthias Hoelzl}
}
\fi

\usepackage[most]{tcolorbox}

\newtcolorbox{colbox}[2][]{%
    tile, 
    colback=#2, 
    size=minimal, left=\oddsidemargin+1in,right=\oddsidemargin+1in, spread sidewards, 
    parbox=false, before upper=\indent, after=\par, #1}

\usepackage{algorithm2e}
\SetKwComment{Comment}{/* }{ */}
\RestyleAlgo{ruled} 
\SetKwInput{kwCompute}{Compute}
\SetKwInput{kwReturn}{Return}

\usepackage{graphicx}

\begin{document}
\maketitle

\begin{abstract}
{\color{black}
One of the most well-established codes for modeling non-linear Magnetohydrodynamics (MHD) for tokamak reactors is JOREK, which solves these equations with a Bézier surface based finite element method. This code produces a highly sparse but also very large linear system. The main solver behind the code uses the'Generalized Minimum Residual Method' (GMRES) with a physics-based preconditioner, but even with the preconditioner there are issues with memory and computation costs and the solver doesn’t always converge well. This work contains the first thorough study of the mathematical properties of the underlying linear system. It enables us to diagnose and pinpoint the cause of hampered convergence. In particular, analyzing the spectral properties of the matrix and the preconditioned system with numerical linear algebra techniques, will open the door to research and investigate more performant solver strategies, such as projection methods.} 
\end{abstract}




\noindent



\section{Introduction}
Extreme global warming and violent geopolitics have made it clear that our society needs new clean energy sources. One approach under active development is controlled nuclear fusion. In this approach, magnetic fields confine a super-hot plasma, mimicking to some extent the conditions that power stars. However, the inherent chaos of plasma dynamics and the high cost of reactor construction underscore the need for robust numerical modeling of plasma dynamics and reactor behavior.

The leading design for controlled fusion is the tokamak, which is a torus lined with superconducting magnets that drive a toroidal magnetic field. Fuseable elements such as Hydrogen and Deuterium are injected into the device and heated until ionization occurs, forming a plasma. The magnetic field then confines these elements within the walls of the device, where at sufficient temperatures and pressures, they can fuse into larger elements and release energy that can be captured.

ITER, the world's most expensive science experiment currently being built, is a 30-meter tall tokamak fusion reactor that aims to demonstrate a significant energy release from the fusion process and spur a generation of commercializable fusion power plants. Its design and analysis require immense computational resources, and one of the codes used to model plasma physics inside ITER is JOREK, developed by an international community including the Max Planck Institute for Plasma Physics (IPP) in Garching, Germany.

JOREK is an advanced and widely-established code used to simulate plasma behavior in tokamaks. It is designed to model plasma instabilities that can shut down a plasma or damage the walls of reactors. JOREK uses a finite element model over Bézier surfaces and a toroidal Fourier decomposition to model toroidal plasmas. It produces a very large system of equations, and has several approaches to solve these efficiently. However, a lot of work still remains to be done to improve the solver efficiency, as the nonlinear model converges slowly.
The purpose of this project is to analyze this system of equations with respect to the physics of the model, and to develop ways to improve solver convergence. Primarily, we intend to build on JOREK's preconditioner, and to develop heuristics by which alternative solvers or alternative models could be suggested to the user.

\begin{figure}[h]
    \centering
    \includegraphics[width=0.4\linewidth]{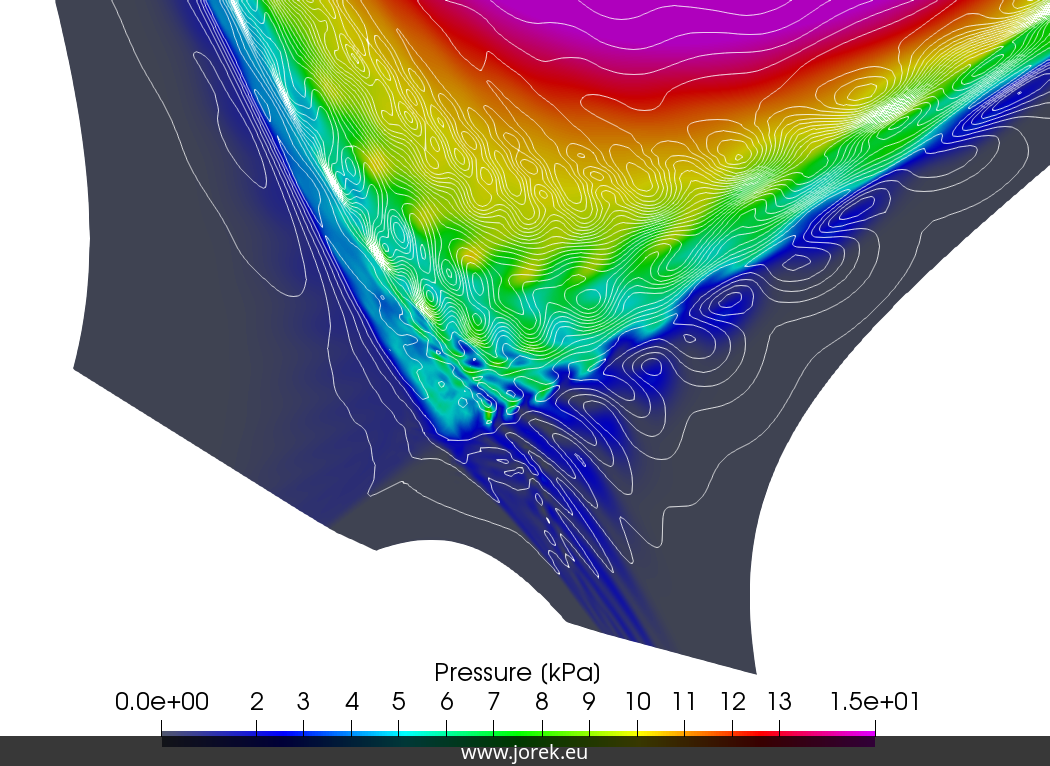}
    \caption{Plasma pressure during an edge-localized-mode in the Joint European Torus plasma.\cite{Futatani_2020}) }
\end{figure}


\section{Plasma Physics and Magnetohydrodynamics}
Plasmas are the most abundant form of matter in the universe. From our sun, to the interstellar medium, lightning and even the glow of neon signs, plasmas are everywhere. Better understanding of them is key to our understanding of all these phenomena. Plasmas are fluids of conducting particles that create and interact with electromagnetic fields. The strong interaction between the plasma and electromagnetic fields means that they are significantly more complicated to understand and model than ordinary liquids and gasses. There are many mathematical models used to study plasmas, but the most well-known of them is Magnetohydrodynamics, often shortened to MHD.

MHD is a set of equations describing the mechanics of electrically conducting fluids such as plasmas or liquid metals. As opposed to particle-based “kinetic” plasma models, MHD is a fluid-flow model that ignores individual particle behavior, instead focusing on the fluid in aggregate. It is similar to the famous Navier-Stokes Equations for fluid flow, but with additional terms for electromagnetic effects. Though less complicated than other plasma models such as Vlasov or the two-fluid model, MHD models are especially well-suited to situations where magnetic forces confine the plasma, such as our usecase in tokamak fusion reactors \cite{Bellan_2006}.

\subsection{Tokamak Dynamics}

The leading candidate for fusion power generation is the tokamak design, which is a magnetic confinement device designed to confine a plasma in the shape of a torus. 
The basic idea behind magnetic-confinement fusion is to confine the charged particles of a plasma at high temperature for the duration necessary to have a high fusion probablility before a particle escapes the system. To achieve this, one applies a strong magnetic field to the plasma. Charged particles traveling in a magnetic field will curve and rotate around magnetic field lines, due to the Lorentz force that acts perpendicular to both the magnetic field and the particle’s direction of motion. The stronger the field, the tighter the radius of these circles (known as the “cyclotron radius”).\cite{Bellan_2006, tokamak_design}. Early plasma confinement devices were linear, but since that does not prevent particles from escaping out of the ends of the device, the idea was to connect the field lines in a torus. For topological reasons, a vector field such as a magnetic field cannot "comb" a sphere flat without points of 0-field at the poles. To prevent particles escaping at those poles, a toroidal design is used.

In a simple toroidal magnetic field, the lines circle the device, which keeps charged particles somewhat confined. Since magnetic field lines always connect back to eachother, the circular geometry of the device means that the magnetic field will be stronger towards the inside of the torus and weaker on the outside. This gradient in the magnetic field causes charged particles to slowly drift laterally, leaving the device. In the tokamak design, currents are added that twist the magnetic field lines helically around the torus, which leads the particle drift to cancel itself out\cite{Wesson_JET}.

JOREK, designed to model toroidal reactors, takes advantage of the toroidal symmetry by formulating its model based on toroidal harmonics. The linear system is divided into block matrices based on toroidal Fourier modes\cite{Hoelzl_2021}.

\subsection{MHD Derivation}
When neglecting certain particle effects and collisionality, plasmas can be described as a fluid. Therefore, to start the derivation, we can start with physical properties of fluids. The easiest place to start is with quantities that will be conserved: mass, momentum, energy, and magnetic flux.

Plasmas conserve mass, so the amount of mass within a volume can only change due to a flux of the mass through the volume. This leads us to the continuity equation:
\begin{equation}
	\partial_t \rho + \nabla \cdot (\rho \mathbf{V}) = 0
\end{equation}
Where $\mathbf{V}$ is fluid flow and $\rho$ is mass density.

Particles in a plasma also conserve momentum, so we adopt the momentum balance equation from Navier Stokes, but with an additional $\mathbf{J} \times \mathbf{B}$ component from the Lorentz force. Here, $p$ is pressure:
\begin{equation}
	\rho \partial_t \mathbf{V} = \mathbf{J} \times \mathbf{B} - \nabla p - \mathbf{V} \cdot \nabla \mathbf{V}
\end{equation}

The current comes from moving current, which itself induces a magnetic field

\begin{equation}
	\rho \mathbf{V} = \mathbf{J} = \frac{1}{\mu_0} \nabla \times \mathbf{B}
\end{equation}

Taking the cross product of the current and the magnetic field yields
\begin{equation}
	\begin{aligned}
		\mathbf{J} \times \mathbf{B} =  \frac{1}{\mu_0} (\nabla \times \mathbf{B}) \times \mathbf{B} \\
 		= -\frac{1}{2\mu_0}\nabla \mathbf{B}^2 + \frac{1}{\mu_0}(\mathbf{B}\cdot \nabla)\mathbf{B}
 	\end{aligned}
\end{equation}

From here, the similarities to fluid equations start to diverge. Charged particles are influenced by the Lorentz force, 

\begin{equation}
    \mathbf{F} = \rho_{e} \mathbf{E} + \rho_{e} \mathbf{V} \times \mathbf{B}    
\end{equation}    

Where $\rho_i$ is the ion charge density, $\rho_e$ is the electron charge density, and $\rho$ is the total charge density. Similarly for pressure:
\begin{equation}    
    \begin{aligned}
        \rho = \rho_{i} - \rho_e \\
        p = p_i + p_e
    \end{aligned}
\end{equation}

From the ideal gas law, pressure relates to density and pressure as $p = \rho T$ with a constant term ($\mu_0$) that drops out due to normalization.

In the stationary case where pressure is balanced by electric current, what follows is a generalized Ohm's law, where $\eta$ is the electrical resistivity \cite{Krebs_Thesis}.

\begin{equation}
    \mathbf{E} + \mathbf{V} \times \mathbf{B} = \eta \mathbf{J} + \frac{1}{en}\left(\mathbf{J}\times\mathbf{B} - \nabla p \right) - \frac{m_e}{e}\frac{d\mathbf{u}}{dt}
\end{equation}

Additional terms come from Maxwell's Equations

\begin{equation}
    \left\{\begin{array}{l}
    	\nabla \cdot \mathbf{E} = \rho / \epsilon_0 \\
    	\nabla \times \mathbf{E} = -\partial_t \mathbf{B} \\
	    \nabla \times \mathbf{B} = \mu_0 \left(\mathbf{J} + \epsilon_0  \partial_t \mathbf{E}\right)\\
	    \nabla \cdot \mathbf{B} = 0
    \end{array}\right.
\end{equation}

One can take certain assumptions if one is only interested in phenomona larger than certain plasma phenomena such as the Debye length (the length scale at which electric fields are screened) and slower than certain frequencies such as the electron cyclotron and plasma frequencies (respectively related to the speed at which electrons orbit magnetic field lines and screen out electric fields). 
For this "Ideal MHD" model, one makes a number of assumptions.
If one assumes that the plasma is quasi-neutral on macroscopic scales and that electrons can quickly displace to balance out charge inequalities, Gauss's Law can be ignored. The resistivity term $\eta$ drops out, as does the whole right side of the generalized Ohm's law, ending up with $\mathbf{E} + \mathbf{V} \times \mathbf{B} = 0$. Since electrons shield electric fields, the $\partial_t \mathbf{E}$ term of Ampere's law disappears.

For normalization, we normalize with respect to the central mass density $\rho_0$, and the vacuum permeability $\mu_0$. Time is normalized time according to the Alfvén time ($\tau_{\mathrm{A}}=a \sqrt{\mu_0 \rho_0} / B_0$), which is the time needed for an Alfvén wave to travel one radian toroidally.

This is a simplified model, and real-world applications need additional extensions, such as finite resistivity, anisotropic heat transport, or two-fluid effects where the electrons and the ions have different properties.

\subsection{Reduced MHD}

Starting from the ideal MHD formulation above, we re-add some resistivity to the current:

\begin{equation}
    \mathbf{E} + \mathbf{V} \times \mathbf{B} = \eta \mathbf{J}
\end{equation}

We then add heat conductivity ($\kappa$), viscosity ($\mu$), diffusivity ($D$), and source terms ($S_H, S_{\rho}$ for heat and particle sources) to the continuity, momentum, and energy equations:

\begin{equation}
\begin{aligned}
& \frac{\partial \rho}{\partial t}+\nabla \cdot(\rho \boldsymbol{v})=\nabla \cdot(D \nabla \rho)+S_\rho, \\
& \rho \frac{d \boldsymbol{v}}{d t}=\boldsymbol{J} \times \boldsymbol{B}-\nabla p+\mu \nabla^2 \boldsymbol{v}, \\
& \frac{d}{d t}\left(\frac{p}{\rho^\gamma}\right)=\nabla \cdot(\kappa \nabla p)+S_H,
\end{aligned}
\end{equation}

The reduced MHD models is a subset of this "resistive and diffusive" MHD model and is designed to reduce the computation costs of the model. It does so by making assumptions that are reasonable for a tokamak configuration. Since it is designed for tokamak plasmas, it uses a cylindrical coordinate system $R, Z, \phi$, along with an associated poloidal coordinate system $r, \theta$.

We start by converting the above equations to cylindrical coordinates, with 

\begin{equation}
    \left\{\begin{array}{l}
    X=R \cos \phi \\
    Y=-R \sin \phi \\
    Z=Z
    \end{array}\right.
\end{equation}

Since $\nabla \cdot \mathbf{B} = 0$, we can break the magnetic field into poloidal and toroidal components $\mathbf{B} = \mathbf{B}_{\phi} + \mathbf{B}_{\theta}$. We make the following assumptions: The magnetic field is dominated by its toroidal component, and the poloidal component is relatively weak ($\mathbf{B}_{\phi} >> \mathbf{B}_{\theta}$). Additionally, the toroidal component of the magnetic field is assumed to be 
constant in time. The ansatz for the magnetic field is:
\begin{equation}
\mathbf{B} = \frac{F_0}{R}\mathbf{e_{\phi}} + \frac{1}{R}\nabla\psi \times \mathbf{e_{\phi}}
\end{equation}
where $F_0$ is a constant, and $\mathbf{e_{\phi}}$ is the toroidal basis vector. $\psi$ is the poloidal magnetic flux.

These assumptions eliminate "fast magnetosonic waves," which are the fastest waves in the system. This allows for larger timesteps as the timestep size depends on the shortest relevant timescales. Additionally, there are fewer unknowns to compute and store\cite{Hoelzl_2021}.

Eventually, (see \cite{Franck_Reduced_MHD} or \cite{Pamela_Thesis} for the full derivation) this comes out to: 

\begin{equation}
\label{eqn:rmhd}
\begin{aligned}
     \frac{\partial \rho}{\partial t} = & -\nabla \cdot(\rho \boldsymbol{v})+\nabla\left(D_{\perp} \nabla_{\perp} \rho\right)+S_\rho \\
     R \nabla \cdot\left[R^2 \rho \nabla_{\perp}\left(\frac{\partial u}{\partial t}\right)\right] = &  {\left[R^4 \rho W, u\right]-\frac{1}{2}\left[R^2 \rho, R^4\left|\nabla_{\perp} u\right|^2\right] } \\
    & -\left[R^2, p\right]+[\psi, j]-\frac{F_0}{R} \frac{\partial j}{\partial \phi}+\mu R \nabla^2 W \\
     \rho F_0^2 \frac{d v_{\|}}{d t} = & F_0 \frac{\partial p}{\partial \phi}-R[\psi, p]+\mu_{\|} \nabla^2 v_{\|} \\
    \rho \frac{\partial T}{\partial t} = & -\rho \boldsymbol{v} \cdot \nabla T-(\gamma-1) p \nabla \cdot \boldsymbol{v}+\nabla \cdot\left(\kappa_{\perp} \nabla_{\perp} T+\kappa_{\|} \nabla_{\|} T\right)+S_T  \\
     \frac{\partial \psi}{\partial t} = & \eta\left(j-j_A\right)+R[\psi, u]-\frac{\partial u}{\partial \phi}
\end{aligned}
\end{equation}

where the velocity $\boldsymbol{v}$ and the toroidal vorticity W , as well as the magnetic field $\mathbf{B}$ and the toroidal current j are defined, respectively, by

\begin{equation}
\begin{aligned}
    & \boldsymbol{v}=\boldsymbol{v}_{\|}+\boldsymbol{v}_{\perp}=v_{\|} \boldsymbol{B}+R^2 \nabla \phi \times \nabla u \\
    & W=\nabla \phi \cdot\left(\nabla \times \boldsymbol{v}_{\perp}\right)=\nabla_{\perp}^2 u \\
    & \boldsymbol{B}=F_0 \nabla \phi+\nabla \psi \times \nabla \phi \\
    & j=-R^2 \nabla \phi \cdot \boldsymbol{J}=\frac{1}{\mu_0} \Delta^* \psi
\end{aligned}
\end{equation}

where $\Delta^*$ is the Grad-Shafranov operator ($\Delta^*\psi = \nabla^2 \psi + 2 \mathbf{B}_\theta$), $u$ is the electric potential, and pressure is defined as $p=\rho T$. Note that the Poisson brackets have been used, with the definition $[a, b]=\boldsymbol{e}_\phi \cdot(\nabla a \times \nabla b)$.




\section{Numerical Discretization}
\label{chap:fem}

Finite Element Methods (FEM) are a technique for solving partial differential equations numerically. The basic idea is to convert an infinitely-dimensional partial differential equation into a linear problem that can be solved with numerical linear algebra techniques.


\subsection{Time Discretization}

Time integrating an equation of the form
\begin{equation}
\frac{\partial \mathbf{A}(\mathbf{u})}{ \partial t}= \mathbf{B}(\mathbf{u}, t)
\end{equation}
can be discretized in general as 

\begin{equation}
\begin{aligned}
& (1+\xi) \mathbf{A}^{n+1}-(1+2 \xi) \mathbf{A}^n+\xi \mathbf{A}^{n-1} \\
& \quad=\Delta t\left[\theta \mathbf{B}^{n+1}+(1-\theta-\phi) \mathbf{B}^n-\phi \mathbf{B}^{n-1}\right]
\end{aligned}
\end{equation}
This is accurate to second order wherever $\phi + \theta - \xi = \frac{1}{2}$. Taking $\phi = 0$ and linearizing (with $\delta\mathbf{u^n} = \mathbf{u^{n+1}} - \mathbf{u^n}$, where $n$ refers to the timestep) gives the equation

\begin{equation}
    \begin{gathered}
    {\left[(1+\xi)\left(\frac{\partial \mathbf{A}}{\partial \mathbf{u}}\right)^n-\Delta t \theta\left(\frac{\partial \mathbf{B}}{\partial \mathbf{u}}\right)^n\right] \delta \mathbf{u}^n} 
    =\Delta t \mathbf{B}^n+\xi\left(\frac{\partial \mathbf{A}}{\partial \mathbf{u}}\right)^n \delta \mathbf{u}^{n-1}
    \end{gathered}    
\end{equation}

For the Crank-Nicolson scheme, $(\theta, \xi) = (1/2, 0)$, for second order BDF2 Gears scheme, $(\theta, \xi) = (1, 1/2)$, and for first order implicit Euler, $(\theta, \xi) = (1, 0)$ \cite{Hoelzl_2021}. After each timestep, we have a rather large linear problem that is then solved.

\subsection{Background and Derivation}
The process underlying FEM is to first cast the problem up into its "weak formulation," which allows us to use linear algebra to solve arbitrary partial differential equations.

Given a system to solve 
\begin{align}
    Au = f \label{linsys}
\end{align}
finding the solution $u \in V$ is equivalent to finding $u \in V$ such that for all "test functions" $v \in V$,
\begin{align}
(Au)(v) = f(v)
\end{align}

Then approximating the weak form of the problem with a finite-dimensional problem by replacing the subspace V of the weak form with a subspace of functions of small, compact, low-degree polynomial "elements" over the domain.
Then selecting a subspace V in $L^2$ and putting the problem in its bilinear Galerkin form \cite{Saad_96, Hughes_FEM, NumII_Lecture}:

From there,
\begin{equation}
\label{eq:weak_eqn}
    \text{Find }u \in V \text{such that } a(u,v) = F(v), \; \forall \; v \in V
\end{equation}

Our computers can only solve finite-dimensional problems, so we perform a dimension reduction:

\begin{equation}
    \text{Find }u_n \in V_n \text{ such that } a(u_n,v_n) = F(v_n), \; \forall \; v_n \in V_n
\end{equation}

This is called the Galerkin equation, and it is a projection of \eqref{eq:weak_eqn} onto $V_n$. From our finite dimensional problem, we now extract a linear system of equations. Since $V_n$ is finite-dimensional, there exists a basis $(\phi_1, ..., \phi_n)$ in $V_n$ that can construct our solution $u_n = = \sum^n_{i=1} \alpha_i \phi_i$.
Due to bilinearity, 
\begin{equation}
    a\left(u_n, v_n\right)=\sum_{i=1}^n \alpha_i a\left(\varphi_i, v_n\right) \stackrel{v_n=\varphi_j}{\Longrightarrow} a\left(u_n, \varphi_j\right)=\sum_{i=1}^n \alpha_i a\left(\varphi_i, \varphi_j\right) .
\end{equation}
This allows us to translate the problem into a linear system we can solve with numerical linear algebra techniques:

\begin{equation}
    A_n \alpha=\left[\begin{array}{ccc}
        a\left(\varphi_1, \varphi_1\right) & \cdots & a\left(\varphi_n, \varphi_1\right) \\
        \vdots & \ddots & \vdots \\
        a\left(\varphi_1, \varphi_n\right) & \cdots & a\left(\varphi_n, \varphi_n\right)
    \end{array}\right]\left[\begin{array}{c}
        \alpha_1 \\
        \vdots \\
        \alpha_n
    \end{array}\right]=\left[\begin{array}{c}
        F\left(\varphi_1\right) \\
        \vdots \\
        F\left(\varphi_n\right)
    \end{array}\right]
\end{equation}

\subsection{Bézier Surfaces}
\label{sec:bezier}

The first version of the JOREK code used "generalized $h-p$ refinable finite elements"\cite{Hoelzl_2021}, but in practice, mesh refinement was impractical. For the second version of JOREK, a new finite element formulation was proposed and implemented that was based on $G^1$ continuous 2D isoparametric cubic Bézier surfaces\cite{Hoelzl_2021}. They are based on interpolating Bernstein polynomials developed by Sergei Bernstein based on the following formulae:

\begin{align}
    \begin{cases}
        B^n_i(s) = C^i_n s^i(1-s)^{n-i}, \\
        C^i_n = \frac{n!}{i!(n-i)!}, \\
        0 \leq i \leq n
    \end{cases}
\end{align}

This gives us a set of polynomials with some useful properties: 
\begin{enumerate}
    \item $B^n_i$ is a basis of $P^n$, the set of polynomials of degree $\leq n$.
    \item $0 \leq B^n_i(s) \leq 1, \forall s \in [0;1]$.
    \item $\sum^n_{i=0} B^n_i(s) = 1, \forall s \in [0;1]$.
\end{enumerate}
In our case, we are using cubic Bézier surfaces: polynomials of degree 3 
\begin{align}
    \begin{cases}
        B^3_0(s) = (1-s)^3, \\
        B^3_1(s) = 3s(1-s)^2, \\
        B^3_2(s) = 3s^2(1-s), \\
        B^3_3(s) = s^3.
    \end{cases}
\end{align}
These polynomials are plotted below:
\begin{center}
    \includegraphics{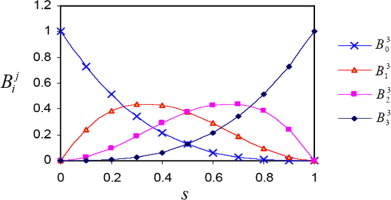} 
    \cite{Czarny_Bezier}
\end{center}

These can be extended into rectangular patches \cite{Czarny_Bezier}, and used as a basis for our FEM model:

\begin{align}
    P(s,t) = \sum^3_{i=0} \sum^3_{j=0} P_{i,j}B^3_i(s)B^3_j(t), \;  0 \leq s, t \leq 1    
\end{align}

Here, $s$ and $t$ are local coordinates where $0 \leq s, t \leq 1$, and $P_{i,j}$ are the 16 control points for the surface. Of these, four correspond to the corners of the patch, and the rest correspond to the tangents ($\frac{\partial P}{\partial s}$, $\frac{\partial P}{\partial t}$) and the cross derivatives ($\frac{\partial^2 P}{\partial t \partial s}$) at the corners.  The patches can be organized (as in JOREK) into an unstructured mesh. In our formulation, Bézier patches are $G^1$ continuous, meaning that where two patches share a common edge, they also share a common angle or tangent at that edge.

Bézier surfaces were chosen for JOREK because of several properties. They require only four degrees of freedom per node, which is an advantage over Lagrangian formulations. They react well to mesh-refinement (unlike pure Hermite formulations or the JOREK I formulation). Bézier surfaces can also be aligned well with the magnetic-fields present, which is advantageous as the physics parallel to the magnetic field differ from the physics perpendicular to the fields \cite{Czarny_Bezier,Krebs_Thesis}.

Bézier surfaces are used to construct the basis for the weak formulation of equations to arrive at the linear system \autoref{linsys}, which is solved using a numerical solution method, which we will discuss in the next section. 



\section{Numerical Solution Methods}


MHD problems produce very large linear systems, due to the multitude of important plasma physics that have effects over many orders-of-magnitude in both time and space. For example, typical fusion plasma dimensions are of the order of a few meters, but the resistive skin depth of the plasma is typically on the order of sub-millimeters, leading to a scale separation of four orders of magnitude \cite{Hoelzl_2021}. This kind of dynamic forces the simulation of large volumes to use a relatively very fine mesh. As a result, this can produce very large linear systems that can be difficult to solve. For the solution to have good performance and stability, a thoughtful application of numerical linear algebra techniques is required.

Solving a system $\mathbf{A}x = b$, where nonsingular $\mathbf{A} \in \mathbb{C}^{n\times n}$, takes on the order of $n^3$ operations if done naively. Taking advantage of certain features of the problem, one can speed this up considerably. With about one in 3000 entries nonzero, the system that JOREK solves is highly sparse. This sparsity makes it a good candidate for certain numerical techniques.

\subsection{Direct Methods}
The basic idea behind direct methods is to decompose the problem into a simpler subproblem. For example, the system \begin{equation}
    \mathbf{A}x=b
\end{equation}
can be decomposed into its lower and upper triangular components 
yielding two simpler problems with triangular matrices that are therefore easy to solve 
\begin{equation}
    \mathbf{L}y=b, \; \mathbf{U}x=y
\end{equation}

A disadvantage with direct methods is that until the algorithm is finished, one does not have any partial solution and the system cannot be analyzed prematurely for information on the state of the solution. They can also have high memory consumption for sparse problems due to fill-in, such as the large sparse systems solved by JOREK. JOREK uses direct solvers only for small problems and to solve blocks of its preconditioner.

\subsection{Iterative Methods}
As opposed to direct methods, iterative methods progressively approximate a solution over numerous iterations. While iterative methods parallelize better, they can fail to converge. Performance can be highly susceptible to the underlying conditioning of the system, often requiring preconditioning. 

\subsubsection{Krylov subspace methods}

Krylov subspace methods are a class of iterative methods that attempt to solve $\mathbf{A}x = b$ by iteratively searching within a limited region of the Krylov subspace of $\mathbf{A}$. 

These are a subclass of so-called projection methods, which look for the approximate solution of the form $x_k = x_0 + \mathcal{S}_k$, where $\mathcal{S}_k$ is a k-dimensional subspace called the "search space." With k degrees of freedom in $x_k$, we also need k constraints. We impose these on the residual ($r_k = b - Ax_k$) with $r_k \perp \mathcal{C}_k$ where $\mathcal{C}_k$ is some space that we have taken as our "constraints space."

Krylov subspace methods use the Krylov subspace as the search space $\mathcal{S}_k = \mathcal{K}_n$. Given an initial guess $x_0$ with an approximate solution $x_m$. Given $\mathbf{A} \in \mathbb{C}^{nxn}$ and $r_0 \in \mathbb{C}^n$, where the residual $r_0 = b-\mathbf{A}x_0$ \cite{Saad_96}, then the Krylov subspace is

\begin{align}
    \mathcal{K}_n(\mathbf{A}, r_0) \coloneqq \text{span}\{r_0, \mathbf{A}r_0, \mathbf{A}^2 r_0,..., \mathbf{A}^{n-1}r_0\}
\end{align}
It can be shown that an iterated Krylov subspace includes an approximate solution\cite{Bai_Krylov}:
Given $x_{x+1} = x_k + \omega(b - \mathbf{A}x_k)$, where $\omega$ is a relaxation parameter and $k=0,1,2,...$, then
\begin{equation}
    x_k = (I - \omega\mathbf{A})x_{k-1} + \omega b =\sum_{j=0}^{k-1}(I-\omega \mathbf{A})^j \omega b \in \mathcal{K}_k(\mathbf{A}, b)
\end{equation}
Provided $x_0 = 0$ and the sequence $x_k$ is convergent, then the solution 
\begin{equation}
x_* \in \mathcal{K}_\infty(\mathbf{A},b)
\end{equation}





JOREK makes use of several different Krylov subspace methods in its solver -- namely Restarted GMRES and BiCGSTAB. 

\subsubsection{Arnoldi Method}

The Arnoldi method is an algorithm for producing an orthonormal basis of vectors. It is used, particularly in other numerical solvers, because it can be made to produce an orthonormal basis of the Krylov subspace. 

Let $\mathbf{A} \in \mathbb{C}^{n \times n}$ and $v \in \mathbb{C}^n \backslash\{0\}$ be of grade $d$ with respect to $\mathbf{A}$. Then there exists $\mathbf{V} \in \mathbb{C}^{n \times d}$ with orthonormal columns and an unreduced upper Hessenberg matrix $\mathbf{H}=\left[h_{i j}\right] \in \mathbb{C}^{d \times d}$, i.e., $h_{i j}=0$ for $i>j+1$ and $h_{i+1, i} \neq 0$ for $i=1, \ldots, d-1$, such that \cite{Liesen_2020}
\begin{align}
    \mathbf{AV} = \mathbf{VH}
\end{align}

The Arnoldi method builds this relation iteratively, via the relation $\mathbf{AV}_m = \mathbf{V}_{m+1}\mathbf{H}_m$, where $\mathbf{V_m} \in \mathbb{C}^{n \times m}, \mathbf{H_m} \in \mathbb{C}^{(m+1) \times m}$. One variant of the algorithm is:

\begin{algorithm}[H]
    \caption{Arnoldi \cite{Saad_96}}\label{alg:arn}
    \kwCompute{Choose a vector $v_1$ of norm 1}
     \For{$j \in 1, 2, ..., m$}{
        \kwCompute{$h_{ij} = (Av_j, v_i) \text{ for } i = 1, 2, ..., j$} 
        \kwCompute{$w_{j} \coloneqq Av_j - \sum_{i=1}^j h_{ij} v_i$}
        $h_{j+1}, j = ||w_j||_2$ \; 
        \If{$h_{j+1}, j 0$}{ Stop }
        $v_{j+1} = w_j/h_{j+1,j}$ \;
    }
\end{algorithm}

If one does not stop the algorithm before the m-th step, then the vectors $v_1, v_2, ..., v_m$ form an orthonormal basis of the Krylov subspace. This forms the basis for many Krylov subspace solvers such as GMRES.

\subsubsection{GMRES}
\label{sec:gmres}

GMRES is an iterative method for nonsymmetric matrices. It uses Arnoldi’s method to compute an orthonormal basis of the Krylov subspace. For the GMRES method, we take constraints space, $\mathcal{C}_m = \mathbf{A}\mathcal{K}_m$, so that $\mathbf{A}\mathcal{K}_m \perp b - \mathbf{A}x_m$\cite{Saad_96}. 

This gives us the following algorithm:

\begin{algorithm}[H]
    \caption{GMRES \cite{Liesen_2020}}\label{alg:gmres}
    \kwCompute{$r_0 = b - Ax_0$}
    \For{$k = 1,2... $}{
        Perform the kth step of Arnoldi to generate $\mathbf{V}_k$ and $\mathbf{H}_{k+1,k}$ \;
        Update the QR factorization of $\mathbf{H}_{k+1,k}$ \;
        Compute the updated residual norm $\left\|r_0\right\|_2\left(\mathbf{Q}_{k+1}^{\mathbf{H}} e_1\right)_{k+1}$ \;        
        \If{the residual norm is small enough}{
            Compute the least squares solution $t_k$ \;
            \kwReturn{the approximate solution $x_k = x_0 + \mathbf{V}_k t_k$}
        }
    }
\end{algorithm}

A major disadvantage of the method is that it needs to store the entire Krylov subspace for every iteration. As the number of iterations grow, the whole subspace must be stored in memory and can become a serious limitation. There are several forms of GMRES that truncate or restart in an attempt to minimize this memory burden \cite{Vuik_Lecture,Ghai_Preconditioned_Krylov}. However, these have their own disadvantages and may not converge as well.

\subsubsection{BiCGSTAB}
\label{sec:bicgstab}

The JOREK team has recently started incorporating the BiCGSTAB algorithm in addition to GMRES. BiCGSTAB produces a residual vector of the form $r_j = \psi_j(\mathbf{A})\phi_j(\mathbf{A})r_0$, where $\psi_j, \phi_j$ are polynomials. $\psi_j$ is defined recursively as $\psi_{j+1}(t) = (1 - \omega_j t) \psi_j(t)$ for some scalar $\omega_j$. BiCGSTAB determines $\omega_j$ by minimizing $||r_j||$ with respect to $\omega_j$.\cite{Ghai_Preconditioned_Krylov,Saad_96}

\begin{algorithm}[h]
    \caption{BiCGSTAB \cite{Saad_96}}\label{alg:bicg}
    \kwCompute{$r_0 = b - \mathbf{A}x_0; r_0^* \mbox{ arbitrary};$}
    Let $p_0 \coloneqq r_0$ \;
    \For{$j = 0, 1, \text{ until convergence }$}{
        $\alpha_j \coloneqq (r_j, r_0^*)/(\mathbf{A}p_j, r_0^*)$ \;
        $s_j \coloneqq r_j - \alpha_j A p_j$ \;
        $\omega_j \coloneqq (\mathbf{A}s_j, s_j)/(\mathbf{A}s_j, \mathbf{A}s_j)$ \;
        $x_{j+1} \coloneqq x_j + \alpha_j p_j + \omega_j s_j$ \;
        $r_{j+1} \coloneqq s_j - \omega_j \mathbf{A} s_j$ \;
        $\beta_j \coloneqq \frac{(r_{j+1},r_0^*)}{(r_j, r_0^*)}\times \frac{\alpha_j}{\omega_j}$ \;\
        $p_{j+1} \coloneqq r_{j+1} + \beta_j (p_j - \omega_j \mathbf{A}p_j)$ \;
    }
\end{algorithm}

BiCGSTAB uses less memory than GMRES, but is less stable and leads to a non monotonic reduction in the residual. It furthermore has an occasional tendency to breakdown or fail to converge. A poor choice of $x_0$ leads to a small $\alpha_i = (r_j, r_0)$ , resulting in small or no improvement per iteration, or even divergence. In that case, one must restart with a new $x_0$ or use another algorithm such as GMRES.



\subsection{Convergence}
{\color{black} Several diagnostic tools are available to numerical analysts to determine convergence properties of the linear systems involved. These depend on properties of the underlying linear system and can be analyzed using numerical linear algebra techniques. 

As a rule of thumb, one often looks at the condition number of an invertible matrix $\mathbf{A}$, which in the $p-$norm is defined 
\begin{equation}
    \kappa(\mathbf{A}) = ||\mathbf{A}||_{p}||\mathbf{A}^{-1}||_{p}.
\end{equation}
In case $\mathbf{A}$ is symmetric and positive definite (SPD), the upper expression in the 2-norm reduces to
\begin{equation}
    \kappa_{2}(\mathbf{A}) = \frac{\lambda_{\max}(\mathbf{A})}{\lambda_{\min}(\mathbf{A})},
\end{equation}
where $\lambda_{\max}$ and $\lambda_{\min}$ denote the respective largest and smallest eigenvalue of the matrix $\mathbf{A}$. A widely propagated misconception is that GMRES convergence can be navigated by looking at the condition number. While this may be true for SPD matrices, this is in general not true for nonnormal $(\mathbf{A}\mathbf{A}^{H} \neq \mathbf{A}^{H}\mathbf{A})$ and indefinite matrices (matrices having both negative and positive eigenvalues). Here, $\mathbf{A}^{H}$ represents the complex conjugate of $\mathbf{A}$.

In general, for normal matrices, the distribution and clustering of the eigenvalues determine the convergence speed of Krylov subspace methods, in particular GMRES. If the eigenvalues are clustered near the point $(1,0)$ in the complex plane, we generally expect fast convergence. For nonnormal matrices however, this may not be the case \cite{Liesen_Tichy,Liesen_2020}. Some numerical evidence has been gathered over the years to suggest that spectral analysis may still provide some notions which could outline convergence behavior \cite{dwarka2020scalable}.

Especially for fusion simulations due to the complexity of the underlying mathematical operators, inadequate conditioning can be misleading in assessing what preconditioning strategies will perform better. Consequently, in this work we focus on unraveling these underlying mathematical properties to interpret the convergence behavior of the current solver in order to work towards acceleration strategies. For example, indefinite nonsymmetric matrices also arise in wave propagation problems, and the respective Krylov based solvers often show acceleration using projection and multigrid techniques (for an overview of examples and literature, see \cite{dwarka2020scalable,dwarka2022scalable}). 

\subsubsection{Preconditioning}

Numerical methods to solve linear systems of equations can be made more efficient by transforming the problem into one that is better solved by that given iterative method\cite{Saad_96}. This process is called "preconditioning," and is extremely important to iterative methods.

For example, given a system $\mathbf{A}x=b$, one wants to find a preconditioner $\mathbf{M}^{-1}$ that is easily invertible and similar to $\mathbf{A}^{-1}$. Then the system $\mathbf{M}^{-1}\mathbf{A}x = \mathbf{M}^{-1}b$ is easy to solve as $\mathbf{M}^{-1}\mathbf{A}$ is close to the identity. \cite{Jorek_Wiki} Of course, finding a matrix close to $\mathbf{A}^{-1}$ is not easy, as finding $\mathbf{A}^{-1}$ is essentially the whole problem to solve. 

In the context of convergence, the preconditioned system $\mathbf{M}^{-1}\mathbf{A}$ should have better spectral properties than the original unpreconditioned system $\mathbf{A}$ in terms of clustering near the point $(1,0)$ in the complex plane. 

\subsubsection{Spectral Analysis}
Unfortunately, getting access to spectral information is computationally very expensive and no analytical expressions for the eigenvalues exist. Consequently, the complexity of the reduced MHD equations require numerical techniques to determine the eigenvalues. 


For the smaller systems, it is computationally feasible to take the full spectrum. However, the larger cases are too large to calculate the complete spectrum. To save computational resources, we use a strategy that enables us to limit our calculations on the important regions of the spectrum.

To calculate the spectrum, we use the "eigsolve" function of the KrylovKit.jl software package \cite{KrylovKit.jl}. This uses the Krylov-Schur algorithm, which uses the Arnoldi method to build a Krylov subspace, see the next subsection. Using the Krylov-Schur method, we are able to take extremal eigenvalues without analyzing the entire system, thus saving computational resources. 

%
Using this method, we take in turns the "largest real", "smallest (most negative) real", "largest imaginary", and "largest magnitude" eigenvalues for each system. We could also get the "smallest (magnitude) imaginary" but this always yields a value with no imaginary component.

Since GMRES convergence is driven by the "radius" of the disk of eigenvalues and minimum distance to the origin, we can infer a complete picture of GMRES convergence performance from just the extremal eigenvalues without taking the complete spectrum. Convergence collapses in GMRES if we have a disk of eigenvalues that includes the origin. Because our preconditioned system consistently yields a spectrum centered at (1,0) in the complex plane, we can develop a relatively effective picture of GMRES convergence by simply taking the most negative real-valued eigenvalue, and seeing if it is negative or comes close to a negative value.}

\subsubsection{Iterative Eigensolver: Krylov-Schur} \label{newseckschur}

  

The Krylov-Schur algorithm is a method used for finding extremal eigenvalues and corresponding eigenvectors of large, sparse matrices. It uses the Rayleigh-Ritz method to find approximate eigenvalue-eigenvector pairs.

In a typically Arnoldi process, the Ritz pairs converge quickly only in an optimal search direction. In practice, many iterations are needed, with more storage and more computation per iteration. To combat this, the algorithm can be restarted in a new initial search direction based on the computed Ritz vectors. A Krylov-Schur decomposition is a special type of Krylov decomposition ($\mathbf{A}\mathbf{V}_m = \mathbf{V}_m\mathbf{B}_m + v_{m+1}b_{m+1}$ but where the matrix $\mathbf{B}_m$ is in "real Schur form," with 1x1 or 2x2 diagonal blocks  \cite{SLEPC-Report}.

The algorithm operates by first forming an orthogonal basis for the Krylov subspace, using the Arnoldi process. Then, it performs a Schur decomposition on the resulting Hessenberg matrix. The resulting Krylov-Schur decomposition is reordered and truncated to order $p$, where $p<m$ is the number of resultant Ritz pairs. The truncation is performed based on how the Ritz values meet a specified convergence criterion. Then the subspace is extended, and the algorithm restarts from the second step. \cite{SLEPC-Report}

\subsection{Preconditioners for MHD Models}

There is some literature investigating preconditioner strategies for different MHD variants or formalisms \cite{Laakmann_Thesis}, but much work remains to be done. Some of the more common preconditioner strategies used for MHD are as follows:

One approach is with multigrid methods, which creates a preconditioner based on solving a coarsened grid. However, these are known to only work well for systems with low Reynold's numbers and coupling \cite{Laakmann_Thesis}.

The most common class of preconditioner used for MHD is the block preconditioner. Where the physics of the problem leads to a system of weakly coupled sub-systems that can be separated into blocks; these blocks can be solved separately for less computational effort than solving the whole larger system directly. If the blocks aren't too strongly coupled, then this is a good approximation of the solution and therefore a good preconditioner. These have been used in incompressible MHD models \cite{Ma_Preconditioners_MHD}, and for the stationary MHD problem, for example\cite{Laakmann_Thesis} by breaking it into hydrodynamic (Navier-Stokes) and electromagnetic (Maxwell's) blocks and then using multigrid to precondition these blocks separately . The JOREK code uses a block preconditioner that will be discussed in detail in \autoref{sec:jorek_precond}.

Another approach is to use an augmented Lagrangian preconditioner. By finding approximate solutions to a weaker and constrained form of the problem that is known as the augmented Lagrangian. This approximate solution can then be used to precondition our original problem. These have found some applicability Navier-Stokes models as well as  incompressible, resistive Hall MHD models, highly coupled and high Reynolds Number plasmas, or anisothermal MHD models \cite{Ma_Preconditioners_MHD,Laakmann_Thesis,Laakmann_2022}.

According to \cite{Laakmann_Thesis}, no practical robust preconditioner yet exists for the problem in general. 

\section{The JOREK Code}

JOREK is a code for the simulation of magnetic confinement fusion reactors. Its goal is to model the dynamics of major plasma disruptions and instabilities, so as to control or minimize them for existing reactors or for new reactors such as ITER\cite{Holod_2021}. It uses the Finite Element Method over Bézier surfaces to model plasma physics using a number of different models, such as full, reduced, or extended MHD equations. To simulate toroidal confinement devices, it uses a toroidal Fourier decomposition\cite{Holod_2021}.

JOREK is written primarily in Fortran 90/95, with some libraries in C and C++ \cite{Holod_2021}. It is massively paralellelized via MPI and OpenMP, and is designed to be run on a high-performance supercomputing cluser such as Marconi-Fusion \cite{Jorek_Wiki}. 





\subsection{Weak Form of Equations}
\label{sec:weak_form}

JOREK takes the MHD equations discussed previously and discretizes them for FEM with Bézier surfaces as the basis. Our test function comes out of our Bézier basis described in \autoref{sec:bezier}: 

\begin{equation}
    \mathcal{T}^*=B_{i, j}(s, t) S_{i, j} e^{i n \phi}
\end{equation}

$B_{i, j}$ is the polynomial Bézier basis, $S_{i, j}$ a scaling factor, $\psi_{i, j, n}$ the coefficients composing the variable $\psi$, and $e^{i n \phi}$ belonging to the Fourier representation, where $n$ refers to the toroidal harmonic\cite{Pamela_Thesis}. For the toroidal basis used in JOREK, the $s$ and $t$ variables are taken to be orthogonal to the $\phi$ basis.

Multiplying our reduced MHD equations (\ref{eqn:rmhd}) by $\mathcal{T}^*$ and integrating over the volume gives us our weak formulation, that we can then discretize.

Using $\mathbf{V}$ as our physical variables and $\mathbf{A}$ is the magnetic vector potential, $\nabla \times \mathbf{A} = \mathbf{B}$, a weak form of the MHD problem is be reformulated as: find $(\mathbf{A}, \mathbf{V}, \rho, p)$ in $\mathcal{V}_{\mathbf{A}} \times \mathcal{V}_{\mathbf{V}} \times \mathcal{V}_\rho \times \mathcal{V}_p$ such that, for any test functions $\left(\mathbf{A}^{\star}, \mathbf{V}^{\star}, \rho^{\star}, p^{\star}\right)$ in $\mathcal{T}_{\mathbf{A}}^* \times \mathcal{T}_{\mathbf{V}}^* \times \mathcal{T}_\rho^* \times \mathcal{T}_p^*$, we have \cite{Hoelzl_2021}:

\begin{equation}
    \begin{split}
    \int \frac{\partial \mathbf{A}}{\partial t} \cdot \mathbf{A}^{\star}dV = & -\int \mathbf{E} \cdot \mathbf{A}^{\star}dV, \\
    \int \rho \frac{\partial \mathbf{V}}{\partial t} \cdot \mathbf{V}^{\star}dV = & -\int(\rho \mathbf{V} \cdot \nabla \mathbf{V}+\nabla p-\mathbf{J})dV \left.\times \mathbf{B}-\nabla \cdot \underline{\tau}-\mathbf{S}_{\mathbf{V}}\right) \cdot \mathbf{V}^{\star}, \\
    \int \frac{\partial \rho}{\partial t} \rho^{\star} = &  -\int\left(\nabla \cdot(\rho \mathbf{V})-\nabla \cdot(\underline{D} \nabla \rho)-S_\rho\right) \rho^{\star} \\
    \int \frac{\partial p}{\partial t} p^{\star} = &  -\int(\mathbf{V} \cdot \nabla p+\gamma p \nabla \cdot \mathbf{V}-\nabla \cdot(\underline{\kappa} \nabla T) \\
    & \; \left.-(\gamma-1) \underline{\tau}: \nabla \mathbf{V}-S_p\right) p^{\star}
    \end{split}
\end{equation}

We transform these into scalar equations by projecting the vectors onto our basis.

\subsubsection{Boundary Conditions}

Both Dirichlet and Neumann boundary conditions can be used for all relevant regions. Where the flux is parallel to the boundary, Dirichlet conditions are assumed. Where the flux intersects the boundary, sheath boundary conditions are used. Certain variables, such as the poloidal flux, current density, electric potential, and vorticity are kept fixed.

The boundary temperatures are constrained by the boundary condition for heat flux. This form assumes that electron and ion temperatures are the same, but these can be separated. $\gamma_{\mathrm{sh}}$ is the sheath transmission factor.

\begin{equation}
\begin{aligned}
\mathbf{q} \cdot \mathbf{n} & \equiv\left(\frac{\rho}{2} \mathbf{V} \cdot \mathbf{V}+\frac{\gamma}{\gamma-1} \rho T\right) \mathbf{V} \cdot \mathbf{n}-\frac{\underline{\kappa}}{\gamma-1} \nabla T \cdot \mathbf{n} \\
& =\gamma_{\mathrm{sh}} \rho T_{\mathrm{e}} \mathbf{V} \cdot \mathbf{n},
\end{aligned}
\end{equation}

Additional boundary conditions given are given, and are expressed in terms of $\underline{\kappa} \nabla T \cdot \mathbf{n}=-\left(c_{\mathrm{b}}-1\right) \rho T \mathbf{V} \cdot \mathbf{n}$:
\begin{equation}
    \left\{\begin{array}{l}
    c_{\mathrm{b}, \mathrm{e}}=(\gamma-1)\left(\gamma_{\mathrm{sh}, \mathrm{e}}-1\right) \\
    c_{\mathrm{b}, \mathrm{i}}=(\gamma-1) \left(\gamma_{\mathrm{sh}, \mathrm{i}}-\gamma-1\right) \\
    c_{\mathrm{b}, \mathrm{total}}=(\gamma-1)\left(\gamma_{\mathrm{sh}} / 2-\gamma / 2-1 \right)
    \end{array}\right.
\end{equation}

\subsection{The JOREK Solver and Preconditioner}
\label{sec:jorek_precond}

For simpler problems that are small or axisymmetric, the system is solved with a direct solver. Usually the PaStiX or STRUMPACK software packages are used, but other solvers are available as well \cite{Hoelzl_2021}.

For more complex systems, JOREK uses a restarted GMRES solver. It must be preconditioned, due to the stiffness and poor-conditioning of the system\cite{Hoelzl_2021}. The preconditioner used is physics and geometry-based, based on the toroidal harmonics that take place in a tokamak structure. 

The matrix problem is written in blocks corresponding to toroidal modes. The preconditioner assumes that toroidal modes are decoupled, so diagonal blocks corresponding to self-interaction are kept while off-diagonal blocks (which correspond to coupling between modes) are dropped \cite{Hoelzl_2021}.

\begin{figure}[h]
    \centering
    \includegraphics{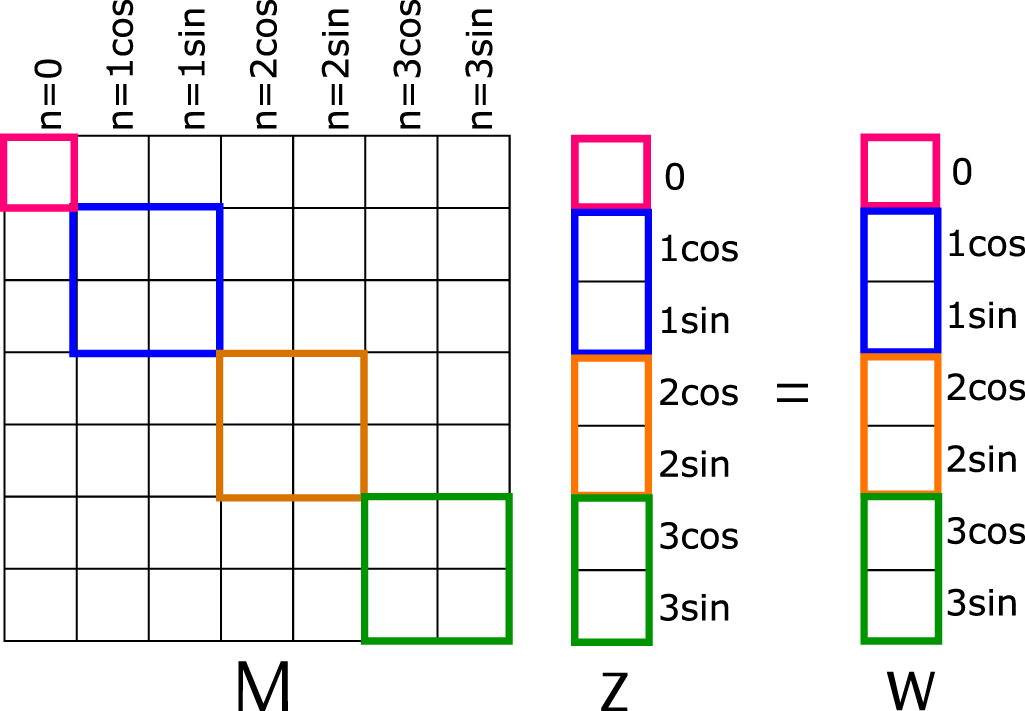} 
    \caption{From \cite{Holod_2021}: "The matrix structure is shown for a simple example case with the toroidal modes $n = (0,1,2,3)$. The color blocks outline the parts of the original matrix A used to form the preconditioner solution $z = M^{-1}\,w$. Each block represent individual toroidal Fourier mode."}
\end{figure}

For linear problems, this is a very effective assumption and the preconditioner behaves well. For highly non-linear problems, the mode-decoupling assumption underlying the preconditioner no longer holds and performance degrades unless a different preconditioner is used.

Current performance problems with the solver are associated with high memory consumption of the factorized preconditioner, poor parallelization of the direct solver used in preconditioning, and poor preconditioner behavior in non-linear cases with strong mode coupling \cite{Hoelzl_2021}. 

\subsubsection{Recent Preconditioner Improvements}

For problems with stronger coupling between modes, a newer preconditioner system is used that breaks the matrix into toroidal "mode groups" that overlap. The preconditioner solves these blocks separately, thus retaining interaction only within these groups. This relaxed assumption requires larger block-matrices to be solved, but the preconditioner matches the true physics of the system more closely and may improve GMRES convergence enough to improve performance overall.

\begin{figure}[h]
    \centering
    \includegraphics{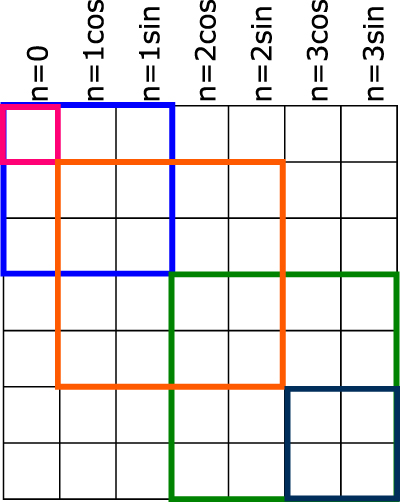}
    \caption{From \cite{Holod_2021}: "Schematic illustration for the Overlapping mode group approach. The color blocks outline the parts of the original matrix A used to form the preconditioner matrices which combine modes (0), (0,1), (1,2), (2,3), and (3) into diagonal blocks."}
\end{figure}


\subsection{Relevant Problem Size}
According to \cite{Hoelzl_2021}, a “typical large problem” leads to a grid size of about 40 million, but the matrix is relatively sparse. This “typical large problem” has “about 12 thousand non-zero entries in each matrix row and about a total of 500 billion non-zero entries, which requires about 4TB of main memory using double precision floating point numbers” \cite{Hoelzl_2021}. This matrix is generally not symmetric, but has symmetric sparsity. At this size, it is too large for a single compute node, so domain decomposition is used to divide the matrix into a decomposed matrix, which is constructed in a parallelized manner.




{\color{red}

}







\section{Numerical Experiments}
For our research, we are going to analyze the convergence properties of example matrices, which represent discretized versions of the reduced MHD model, that have been generated under various JOREK settings. We were given access to a standalone JOREK solver that was written in Fortran90 and C++ as a reference and for benchmarking, but the analysis in this work was done using Julia due to its performance in numerical computing, its readability and relative ease-of-use. 


\subsubsection*{Simple Tearing Mode Case in Limiter Geometry}
The first problem analyzed has a simple toroidal geometry with a circular cross-section without an X-point (“limiter geometry”) with the major radius of the torus being 10 times higher than the minor radius of the circular cross section (“large aspect ratio”). As physics model, the reduced visco-resistive MHD model of JOREK without parallel velocity is used. The anisotropy of the heat conduction is low. The pressure of the plasma is so low that it does not contribute to the dynamics (“low beta”). This plasma develops a slowly growing so-called tearing mode instability dominated by the toroidal mode number n=1 that is destabilized by the radial profile of the plasma current and leads to the reconnection of magnetic field lines and the formation of magnetic islands. The toroidal Fourier spectrum used to model the case includes only three toroidal modes.


For this model, we were given an h5 file describing a sparse square matrix of n=20,160, with 21,081,600 nonzero elements, and 3 preconditioner blocks corresponding to the three different toroidal modes n=0,1,2. As discussed in (\autoref{sec:jorek_precond}), the first preconditioner block corresponds to the first toroidal mode, while subsequent blocks are of duplicate dimension, since they contain cosine and sine components. 
The first toroidal mode block is a sparse square matrix of n=4,032 with 843,264 nonzero elements. The following two blocks both have n=8,064 with 3,373,056 nonzero elements.




\subsubsection*{Ballooning Mode Case in X-Point Geometry}
The second cases we analyzed is considerably more realistic. It contains a plasma with an aspect ratio (major radius divided by the minor radius) that is typical for tokamak experiments, has an X-point, a plasma pressure that influences the dynamics, and a more realistic heat diffusion anisotropy. Furthermore, the visco-resistive reduced MHD model of JOREK including flows along magnetic field lines is used. Driven by the radial gradient of the pressure, the plasma develops a so-called ballooning mode instability with higher toroidal mode numbers than the previous case, a type of interchange instability. 

We were given three different variants of this case by the JOREK team: one artificially small case (44,667 rows/columns) with two toroidal modes $n=0$ and 6, a similar case with higher FEM resolution (350,679 rows/columns), and another even larger case (584,465 rows/columns) that has three toroidal modes $n=0, 6, 12$. See \autoref{table:303_cases} for more details.


\begin{table}[h]
    \centering
    \scalebox{0.92}{
    \begin{tabular}{|c|c|c|c|c|}
        \hline
        Case &  \# of Toroidal Modes & n & Nonzero Elements & Toroidal Mode Size \\ \hline
        "Small"   & 3   & 44,667   & 32,219,901 & 14,889  \\ \hline
        "Medium"   & 3   & 350,679   & 260,359,785  & 116,893 \\ \hline
        "Large"   & 5  & 584,465  & 723,221,625 & 116,893 \\ \hline
    \end{tabular}}
    \caption{System Sizes for the Ballooning-mode Cases. Toroidal Mode Size refers to the matrix size of our first toroidal mode block, and is double the size for subsequent blocks.}
    \label{table:303_cases}
\end{table}


\subsection{Initialization}
For the GMRES and BiCGSTAB implementations, we used the implementations included in the Krylov.jl library \cite{Krylov.jl}. For spectral decompositions or large systems, we used the eigsolve function from the KrylovKit.jl library \cite{KrylovKit.jl}. This uses an Arnoldi iteration method called Krylov-Schur to find extremal eigenvalues, see \autoref{newseckschur}.

\subsubsection{Hardware}
For our analysis, we used the Marconi supercomputing cluster. The Marconi supercomputer has 3,188 nodes, with 2 24-core Intel Xeon 8160 (SkyLake) processors and 196 GB of ram per node \cite{Marconi}. We only executed on one node at a time for our analysis, and made use of only 3-5 processors per node, depending on the system's construction.

\subsubsection{Preconditioner Operator}
\label{sec:pc_operator}
It should be noted that the JOREK solver does not use an explicit preconditioner matrix, but instead preconditions by solving the LU decompositions of the preconditioner blocks against a solution vector. We utilize an "Operator" object and attached multiplication operations to it which behave identical to our preconditioning process. 


For our analysis, we constructed an Operator out of our preconditioning algorithm and used that as though it was an explicit $M^{-1}$ matrix. We verified that this object had identical solver behavior compared to our explicit preconditioner matrix.


%


\subsection{Simple Tearing Mode Case in Limiter Geometry}
\label{chap:199}

\subsubsection{Solver Behavior}

\begin{figure}[h]
    \centering
    \begin{minipage}{.5\textwidth}
        \centering
        \includegraphics[width=\linewidth]{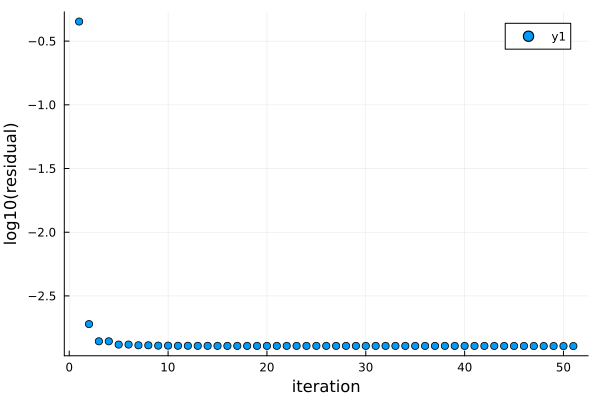}
        \caption{Residual per Iteration for GMRES for the Non-Preconditioned Problem}
        \label{fig:199_gmres_raw}
    \end{minipage}%
    \begin{minipage}{.5\textwidth}
        \centering
        \includegraphics[width=\linewidth]{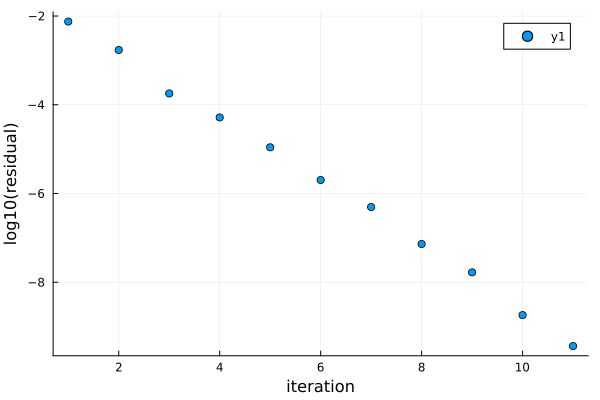}
        \caption{Residual per Iteration for GMRES for the Preconditioned Problem}
        \label{fig:199_gmres_pc}
    \end{minipage}
\end{figure}

Running GMRES without preconditioning, we get very poor convergence, as can be seen in \autoref{fig:199_gmres_raw}. The residual norm drops in the first two iterations, but then remains relatively flat.

With the block preconditioner described in (\autoref{sec:jorek_precond}), we have convergence to a residula of $10^{-9}$ in 10 iterations (\autoref{fig:199_gmres_pc}). The residual norm drops exponentially.

\begin{figure}[h]
    \centering
    \begin{minipage}{.5\textwidth}
        \centering
        \includegraphics[width=\linewidth]{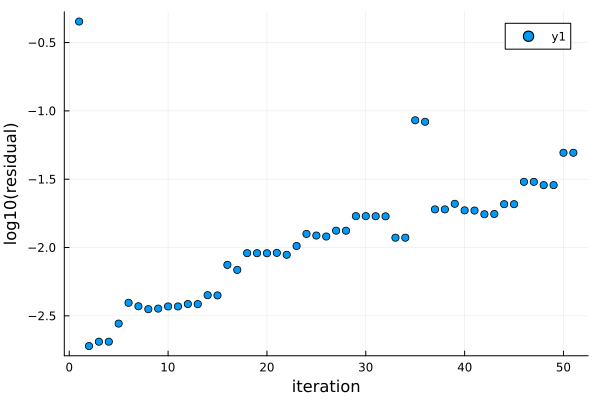}
        \caption{Residual per Iteration for BiCGSTAB for the Non-Preconditioned Problem}
        \label{fig:199_bicgstab_raw}
    \end{minipage}%
    \begin{minipage}{.5\textwidth}
        \centering
        \includegraphics[width=\linewidth]{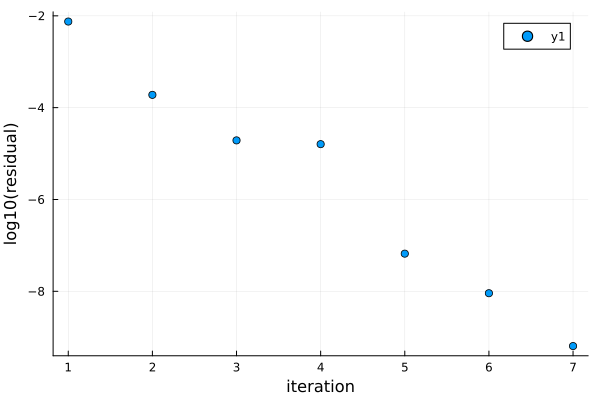}
        \caption{Residual per Iteration for GMRES for the Non-Preconditioned Problem}
        \label{fig:199_bicgstab_pc}
    \end{minipage}
\end{figure}

Similar behavior is seen with BiCGSTAB. Without preconditioning, the residual norm actually increases slightly (\autoref{fig:199_bicgstab_raw}. With preconditioning, the residual norm drops exponentially to $10^{-9}$ after only 7 iterations (\autoref{fig:199_bicgstab_pc}).

\subsubsection{Spectral Analysis}
Calculating the eigenvalues for the unpreconditioned problem, we see a wide range in the complex plane, stretching from approximately $-10^5$ to $10^{12}$ on the real number line, with values as small as $\pm 5 \cdot 10^{-7}$, and as large as $\pm 140i$ in the imaginary. 

This is an indefinite, and extremely poorly conditioned system. It is exactly the kind of system that would have poor performance in GMRES. It is real-valued, but not symmetric, so it does not lend itself to other more performant methods such as CG.

After preconditioning however, we get a spectrum centered at $1$ with values ranging  $\pm 0.2$ in the real and $\pm 0.35i$ in the imaginary. As a system with tightly clustered eigenvalues, this spectrum explains the effectiveness of the preconditioning system used by JOREK. An effective preconditioner (one that roughly emulates $\mathbf{A}^{-1}$) leads to a system centered at 1 on the complex plane, with all values tightly clustered around 1 and all values greater than 0.

\begin{figure}[h]
    \centering
    \begin{minipage}{.5\textwidth}
        \centering
        \includegraphics[width=\linewidth]{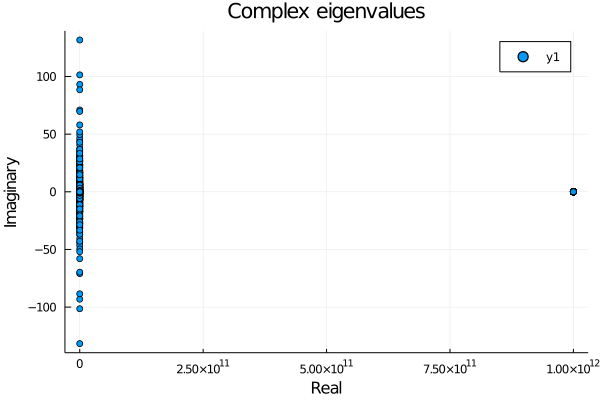}
        \caption{Spectrum for the non-preconditioned matrix $\mathbf{A}$}
        \label{fig:image1}
    \end{minipage}%
    \begin{minipage}{.5\textwidth}
        \centering
        \includegraphics[width=\linewidth]{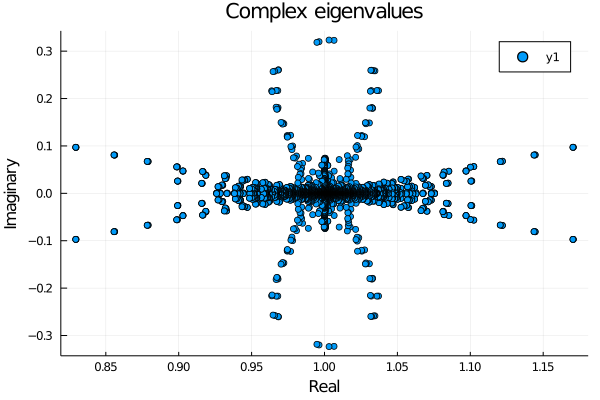}
        \caption{Spectrum for the preconditioned matrix $\mathbf{M}^{-1}\mathbf{A}$}
        \label{fig:image2}
    \end{minipage}
\end{figure}



\subsection{Ballooning Mode Case in X-Point Geometry}
\label{chap:303}

\subsubsection{Solver Behavior}

Similarly to the tearing mode case, convergence without preconditioning was poor. Using the simple block preconditioner however, the solution converges well. The "small" case reached a residual of $10^{-9}$ in 16 iterations. 

\begin{figure}[!htb]
    \centering
    \begin{minipage}{.5\textwidth}
        \centering
        \includegraphics[width=\linewidth]{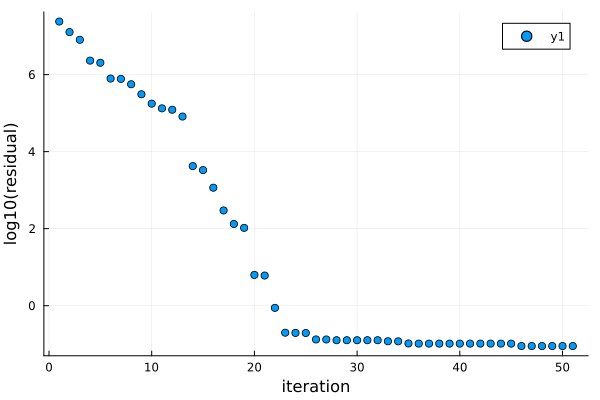}
        \caption{Residual per Iteration for GMRES for the Non-Preconditioned "Small" Problem}
        \label{fig:303_small_gmres_raw}
    \end{minipage}%
    \begin{minipage}{.5\textwidth}
        \centering
        \includegraphics[width=\linewidth]{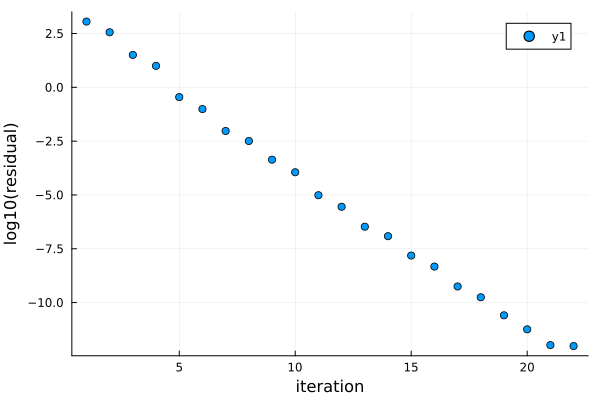}
        \caption{Residual per Iteration for GMRES for the Preconditioned "Small" Problem}
        \label{fig:303_small_gmres_pcd}
    \end{minipage}
\end{figure}

\begin{figure}[!htb]
    \centering
    \begin{minipage}{.5\textwidth}
        \centering
        \includegraphics[width=\linewidth]{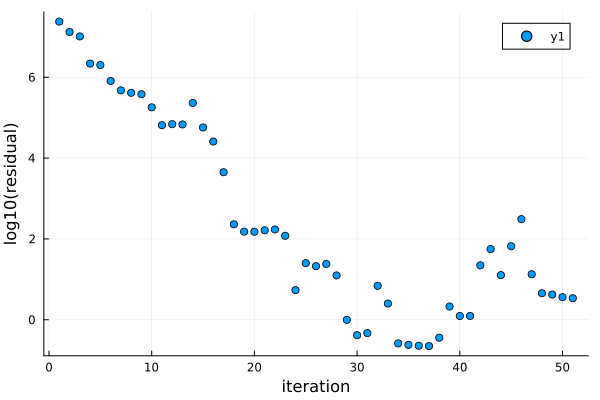}
        \caption{Residual per Iteration for BiCGSTAB for the Non-Preconditioned Problem}
        
    \end{minipage}%
    \begin{minipage}{.5\textwidth}
        \centering
        \includegraphics[width=\linewidth]{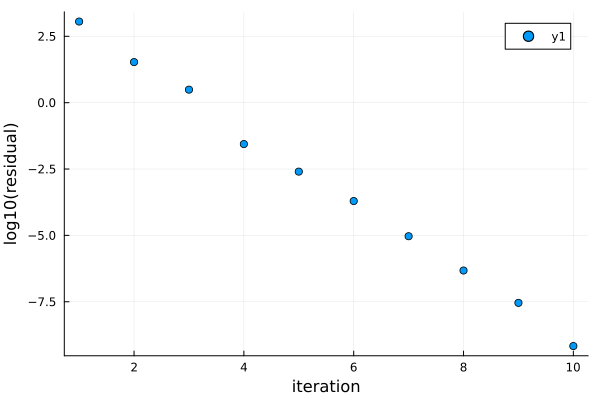}
        \caption{Residual per Iteration for BiCGSTAB for the Preconditioned Problem}
        
    \end{minipage}
\end{figure}

\begin{figure}[!htb]
    \centering
    \begin{minipage}{.5\textwidth}
        \centering
        \includegraphics[width=\linewidth]{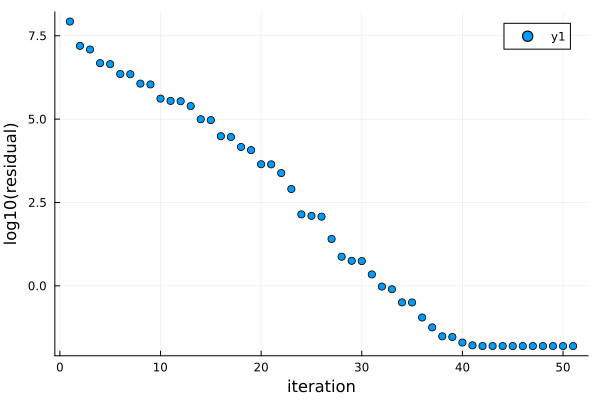}
        \caption{Residual per Iteration for GMRES for the Non-Preconditioned "Medium" Problem}
        \label{fig:303_medium_gmres_raw}
    \end{minipage}%
    \begin{minipage}{.5\textwidth}
        \centering
        \includegraphics[width=\linewidth]{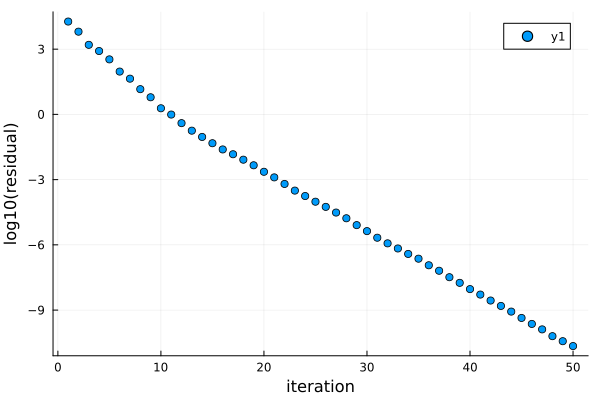}
        \caption{Residual per Iteration for GMRES for the Preconditioned "Medium" Problem.}
        \label{fig:303_medium_gmres_pcd}
    \end{minipage}
\end{figure}

\begin{figure}[!htb]
    \centering
    \begin{minipage}{.5\textwidth}
        \centering
        \includegraphics[width=\linewidth]{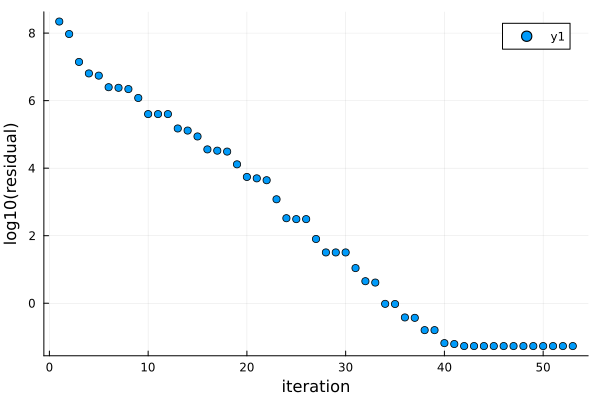}
        \caption{Residual per Iteration for GMRES for the Non-Preconditioned "Large" Problem}
        \label{fig:303_large_gmres_raw}
    \end{minipage}%
    \begin{minipage}{.5\textwidth}
        \centering
        \includegraphics[width=\linewidth]{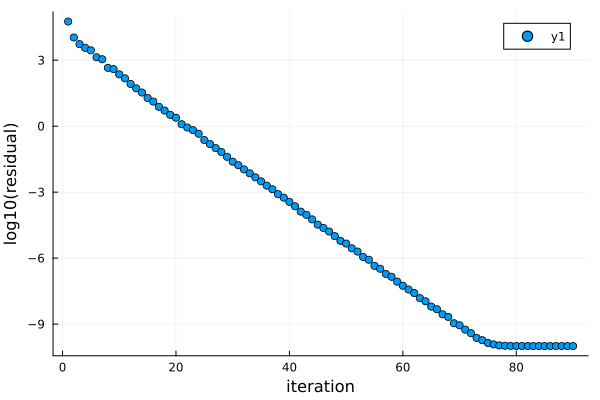}
        \caption{Residual per Iteration for GMRES for the Preconditioned "Large" Problem.}
        \label{fig:303_large_gmres_pcd}
    \end{minipage}
\end{figure}

\subsubsection{Spectral Analysis}

For the smaller ballooning mode system, the spectrum shows similar patterns as for the tearing mode case. The un-preconditioned state is highly indefinite, with eigenvalues tracking in the reals from -3.9e10 to 1e12, and with eigenvalues as low as 1e-7

Applying the preconditioner matrix, we now see a spectrum centered around $1 \pm (0.2 + 0.35i)$. These two spectra explain (as they did for the tearing mode case) the convergence behavior for GMRES in the preconditioned and un-preconditioned forms.

\begin{figure}[!htb]
    \centering
    \begin{minipage}{.5\textwidth}
        \centering
        \includegraphics[width=\linewidth]{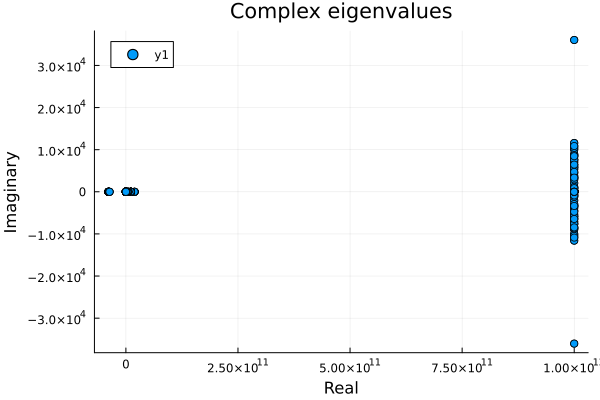}
        \caption{Complex spectrum for A in the small ballooning-mode case}
    \end{minipage}%
    \begin{minipage}{.5\textwidth}
        \centering
        \includegraphics[width=\linewidth]{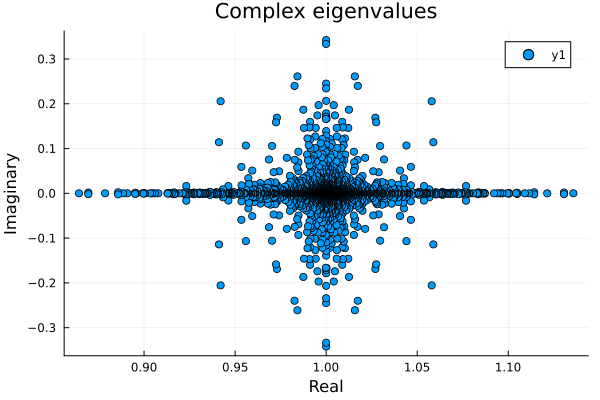}
        \caption{Preconditioned spectrum $M^{-1}A$ for the small ballooning-mode case}
        
    \end{minipage}
\end{figure}

We were only able to fully analyze the smaller of the ballooning-mode systems. The larger systems, are too large to take the full spectrum, but an incomplete spectrum can still tell us about convergence behavior.

Using the Krylov-Schur method, we are able to take extremal eigenvalues without analyzing the entire system, saving lots of computation. We used the KrylovKit.jl Krylov-Schur "eigsolve" method to take in turns the "largest real", "smallest real", "largest imaginary", "smallest imaginary", and "largest magnitude" eigenvalues for each system. 

Since GMRES convergence is driven by the "radius" of the disk of eigenvalues and minimum distance to the origin, we can infer a complete picture of GMRES convergence performance from just the extremal eigenvalues.

As can be seen below for the medium size case, the spectrum is again centered around 1 in a disk that does not include the origin. However, the radius is larger than for the smaller case, which helps to explain the reduced GMRES convergence.

\begin{figure}[!htb]
    \centering
    \begin{minipage}{.5\textwidth}
        \centering
        \includegraphics[width=\linewidth]{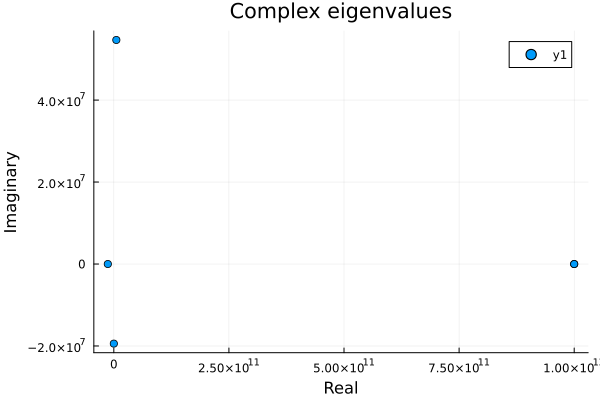}
        \caption{Extremal eigenvalues  for $\mathbf{A}$ in the medium ballooning-mode case}
    \end{minipage}%
    \begin{minipage}{.5\textwidth}
        \centering
        \includegraphics[width=\linewidth]{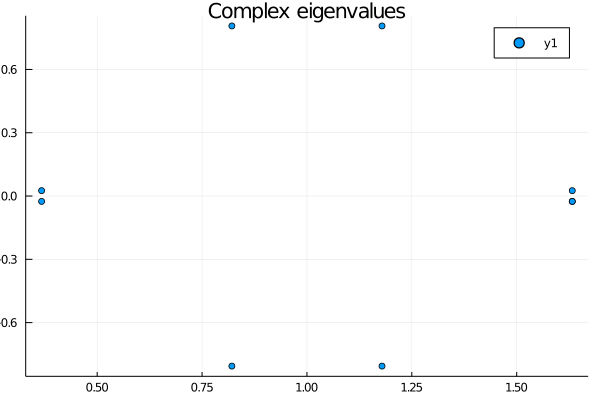}
        \caption{Extremal eigenvalues for the preconditioned spectrum $\mathbf{M}^{-1}\mathbf{A}$ for the medium ballooning-mode case}
        
    \end{minipage}
\end{figure}

\begin{figure}[H]
    \centering
    \begin{minipage}{.5\textwidth}
        \centering
        \includegraphics[width=\linewidth]{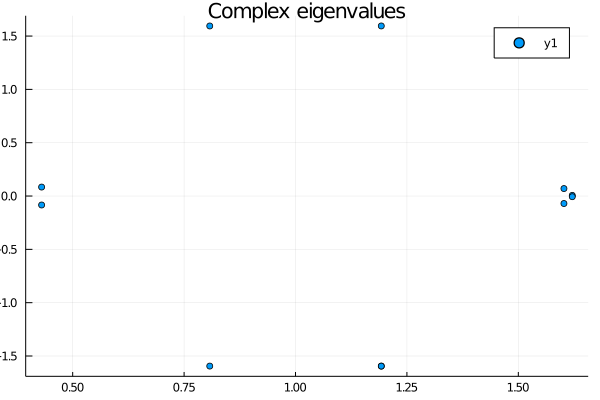}
        \caption{Extremal eigenvalues for the preconditioned spectrum $\mathbf{M}^{-1}\mathbf{A}$ for the large ballooning-mode case}
        
    \end{minipage}
\end{figure}

We obtain a very similar plot for the larger case. For a full table of the range of eigenvalues in the complex domain, see \autoref{table:pcd_303_evalues}

\begin{table}[!htb]
    \centering
    \begin{tabular}{|c|c|c|c|c|}
        \hline
        Case & Smallest Real & Largest Real &  Largest Imaginary \\ \hline
        "Small"   & -3.90e10   & 1.00e12  & $1.00e12 \pm 6.75e4i$ \\ \hline
        "Medium"  &  -1.3e10   & 1.00e12   &  $1.00e12 \pm 2.8e4i$ \\ \hline
        "Large" & -1.30e10  & 1.00e12  &   $1.00e12 \pm 2.8e4i$ \\ \hline
    \end{tabular}
    \caption{Spectral properties for the ballooning-mode cases without preconditioning.}
    \label{table:raw_303_evalues}
\end{table}

\begin{table}[!htb]
    \centering
    \scalebox{0.8}{
    \begin{tabular}{|c|c|c|c|c|c|}
        \hline
        Case & GMRES Iterations (JOREK/Julia) & Smallest Real & Largest Real &  Largest Imaginary \\ \hline
        "Small"   & NA/16  & $0.86$  & $1.14 + 0.0i$ & $1.06 + 0.25im$ \\ \hline
        "Medium"  & 23/38 & $0.37 \pm 0.025i$ & $1.63 \pm 0.025i$   &  $1.18 + 0.81i$ \\ \hline
        "Large" & 40/59  & $0.39 \pm 0.61i$  & $1.65 \pm 0.56i$  & $1.13 + 1.40i$ \\ \hline
    \end{tabular}}
    \caption{Preconditioned solution behavior and spectral properties for the ballooning-mode cases.}
    \label{table:pcd_303_evalues}
\end{table}

\subsection{Parameter Studies}

The JOREK team has observed that several model parameters and settings can severely impact convergence, so they asked us to look at models for a few different systems. These new systems are otherwise identical to the "large" ballooning-mode system, which we can use here as a reference.

\subsubsection{Stale Preconditioner}
The JOREK team has observed that several model parameters and settings can severely impact convergence, so they asked us to look at models for a few different systems. These new systems are otherwise identical to the "large" ballooning-mode system, which we can use here as a reference.

\subsection{Stale Preconditioner}

As the system evolves, the JOREK uses the same preconditioner since the LU decompositions required to update the preconditioner are computationally expensive. Over successive iterations, the effectiveness of the preconditioner drifts and GMRES convergence collapses.

\begin{figure}[H]
    \centering
    \begin{minipage}{.5\textwidth}
        \centering
        \includegraphics[width=\linewidth]{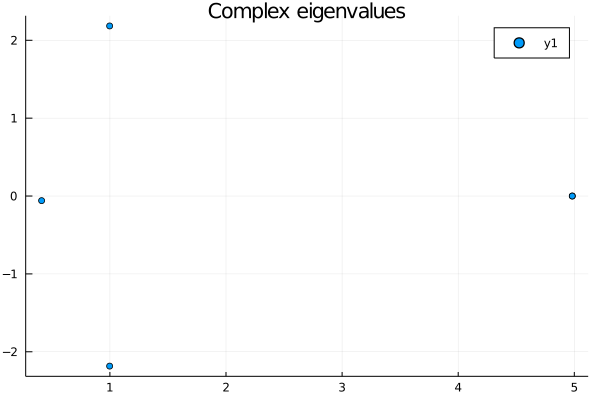}
        \caption{Extremal eigenvalues for the preconditioned spectrum $M^{-1}A$ for the large ballooning-mode case with a stale preconditioner.}
        \label{fig:stale_eigs}
    \end{minipage}%
    \begin{minipage}{.5\textwidth}
        \centering
        \includegraphics[width=\linewidth]{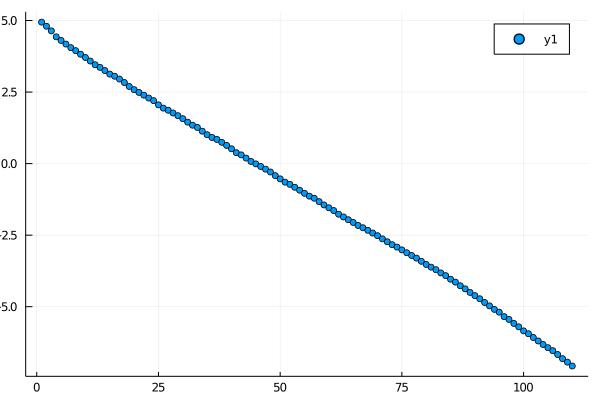}
        \caption{Residual per Iteration for GMRES for the "stale" Preconditioned Problem.}
        \label{fig:stale_gmres}
    \end{minipage}
\end{figure}

We ran an experiment on a preconditioner that is 5 time-steps out-of-date. It had much worse convergence behavior (see \autoref{fig:stale_gmres}) and its spectrum was considerably worse (see \autoref{fig:stale_eigs} and \autoref{table:experiment_eigs}). The stale preconditioner is no longer a good approximation of the system. In the original preconditioned system, $\mathbf{M}^{-1}$ is the inverse of $\mathbf{A}$ with some assumptions about mode coupling. After several time-steps, one has evolved, and $\mathbf{M}$ no longer really approximates $\mathbf{A}$, and $\mathbf{M}^{-1}\mathbf{A}$ no longer approximates the identity. This leads to a divergence in the spectrum and a consequent impact on the convergence for GMRES. In the case of our 5 time-steps stale preconditioner, the spectrum has grown from a width of 1.26 in the reals to a width of 4.57, resulting in a much larger {\color{black}spectral "radius"}.

\subsubsection{Parallel Thermal Conductivity}

As the JOREK team believes the parallel thermal conductivity parameter (known as $\kappa_{par}$) could be a driver for poor convergence, we also ran a series of experiments with different values for $\kappa_{par}$, the parallel thermal conductivity. Due to the way charged particles move in magnetic fields, the thermal conductivity in a tokamak plasma is extremely anisotropic. Parallel thermal conductivity along magnetic field lines is much larger than perpendicular thermal conductivity, that moves across magnetic field lines.

The JOREK team supplied us with two additional runs with $\kappa_{par}$ set respectively at 100 and 200. The systems are otherwise identical to the "large" system, so we will use that system as our reference.

As can be seen in the following figures and table, the convergence degrades considerably compared to our reference system. For both of these systems, the smallest eigenvalue is considerably closer to 0, which is where we would expect to see a complete collapse of GMRES convergence. This explains why we see such a large change in our required GMRES iterations, from 59 in the reference to 81 for $\kappa_{par}=100$.

The difference in convergence and spectrum between the $\kappa_{par}$ 100 and 200 systems though is fairly small. As expected, the $\kappa_{par}=200$ system has slightly larger spectral radius, and slightly worse convergence behavior. However, the difference is minimal, suggesting that larger values of $\kappa_{par}$ only make a significant difference up to a certain value.

\begin{figure}[H]
    \centering
    \begin{minipage}{.5\textwidth}
        \centering
        \includegraphics[width=\linewidth]{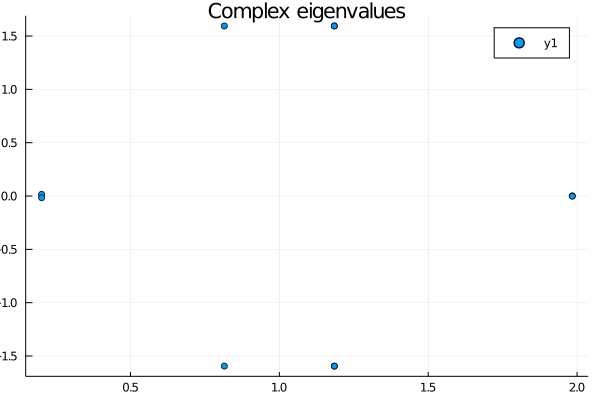}
        \caption{Extremal eigenvalues for the preconditioned spectrum $M^{-1}A$ for the large ballooning-mode case with a stale preconditioner.}
    \end{minipage}%
    \begin{minipage}{.5\textwidth}
        \centering
        \includegraphics[width=\linewidth]{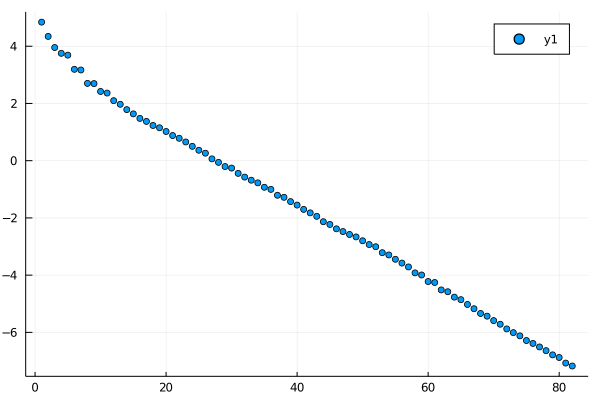}
        \caption{Residual per Iteration for GMRES for the Preconditioned Problem with $\kappa_{par}=100$}
        
    \end{minipage}
\end{figure}


\begin{figure}[H]
    \centering
    \begin{minipage}{.5\textwidth}
        \centering
        \includegraphics[width=\linewidth]{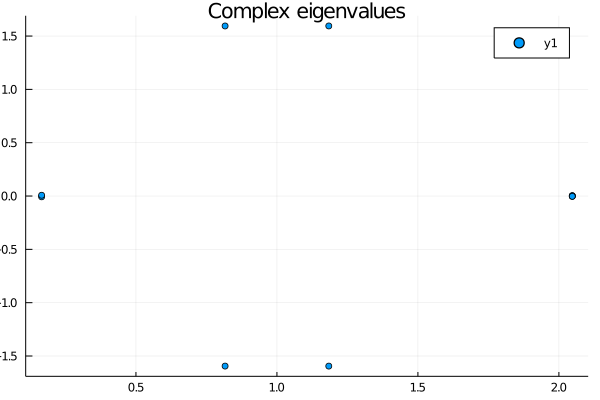}
        \caption{Extremal eigenvalues for the preconditioned spectrum $\mathbf{M}^{-1}\mathbf{A}$ for the large ballooning-mode case with $\kappa_{par}=200$}
    \end{minipage}%
    \begin{minipage}{.5\textwidth}
        \centering
        \includegraphics[width=\linewidth]{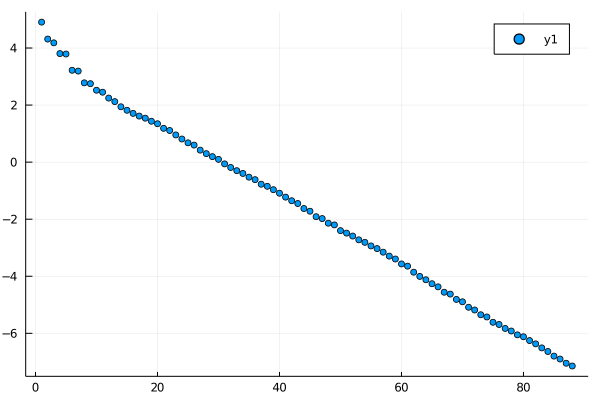}
        \caption{Residual per Iteration for GMRES for the $\kappa_{par}=200$ Preconditioned Problem.}
        
    \end{minipage}
\end{figure}
\begin{table}[H]
    \centering
    \scalebox{0.78}{
    \begin{tabular}{|c|c|c|c|c|c|}
        \hline
        Case & GMRES Iterations & Smallest Real & Largest Real &  Largest Imaginary \\ \hline
        Reference   & 59  & $0.39 \pm 0.61i$  & $1.65 \pm 0.56i$  & $1.13 + 1.40i$ \\ \hline
        Stale Preconditioner  & 109 & $0.412 \pm 5.9e-2i$ & $4.98 \pm 4.18e-9i$   &  $1.00 \pm 2.18i$ \\ \hline
        $\kappa_{par}=100$ & 81  & $0.20 \pm 1.57e-2i$  & $1.98 
    \pm 1.81e-3i$  & $1.18 \pm 1.59i$ \\ \hline
        $\kappa_{par}=200$ & 87  & $0.165 \pm 7.39e-3i$  & $2.05 \pm 2.76e-3i$  & $1.18 \pm 1.59i$ \\ \hline
    \end{tabular}}
    \caption{Preconditioned solution behavior and spectral properties for additional the ballooning-mode cases.}
    \label{table:experiment_eigs}
\end{table}


As our systems get less ideal, the  "tails" of our spectrum grow along the real number line, reaching towards the y-axis on the left and higher values on the right. As they grow, they not only increase the spectral radius, but as the "tail" grows on the left towards 0, it also reduces the spectral disk's distance from the origin, influencing GMRES convergence. This relation to convergence can be seen in the growing number of GMRES iterations necessary to reach our required tolerance.


\section{Conclusions and Future Work}
{\color{black}
In this work we present the first spectral analysis from a numerical analysis point of view to thoroughly understand the convergence behavior of solvers used in fusion simulations. Discretized versions of the reduced MHD equations often lead to sparse, complex, indefinite and nonsymmetric systems. Consequently, the choice of numerical solvers is nontrivial as state-of-the-art solvers do not apply to these type of matrices. As a result, convergence can be slow and difficult to improve. By studying the underlying mathematical properties of the matrices, such as eigenvalues, we have diagnostic tools to interpret the convergence and find solutions to accelerate the simulation speed. 
The spectrum observed in these examples reiterate a classical point of view, often encountered in numerical analysis: as the model problems become more involved, the underlying eigenvalues start growing tails, leading to hampered convergence.
Now that we have established this effect, we can use a set of intricate techniques to enhance the preconditioner, such as subspace projection methods, which will addressed in future work.}











\bibliographystyle{plain}

\end{document}